\documentclass[prb,twocolumn,showpacs,preprintnumbers,eqsecnum]{revtex4}
\usepackage{graphicx}
\usepackage{dcolumn}

\begin{document}

\title{Effects of Electron-Electron and Electron-Phonon Interactions
 \\ in Weakly Disordered Conductors and Heterostructures}


\author{ A. Sergeev}
 \email{sergeev@ciao.eng.wayne.edu}
\affiliation{Department of ECE, Wayne State University, Detroit,
Michigan 48202}
\author{M. Yu. Reizer} \affiliation{5614 Naiche Rd.
Columbus, Ohio 43213}
\author{V. Mitin} \affiliation{Electrical Engineering
Department, University at Buffalo, Buffalo, New York 14260}

\begin{abstract}
We investigate quantum corrections to the conductivity due to the
interference of electron-electron (electron-phonon) scattering and
elastic electron scattering from impurities and defects in weakly
disordered conductors. The interference corrections are
proportional to the Drude conductivity and have various
temperature dependencies. The electron-electron interaction
results in the $T^2 \ln T$-correction in a bulk conductors. In a
quasi-two-dimensional conductor, $d <L_T = v_F/T$ ($d$ is the
thickness, $v_F$ is the Fermi velocity), with 3D electron spectrum
($p_F d>1$) this correction is linear in temperature and differs
from that for 2D electrons (G. Zala et. al., Phys. Rev.B {\bf 64},
214204 (2001)) by a numerical factor. In quasi-one-dimensional
conductors with 3D and 2D electron spectra (a wire with radius
$r<L_T$ and a strip with width $b<L_T$), temperature-dependent
corrections are proportional to $\ln T$. The value and the sign of
the corrections depend on the strength of the electron-electron
interaction in the triplet channel. The electron interaction via
exchange of virtual phonons gives $T^2 \ln T $-correction. In bulk
semiconductors the interaction of electrons with thermal phonons
via the screened deformation potential results in $T^6$-term and
via unscreened deformation potential leads to $T^2$-term. For a
two-dimensional electron gas in heterostructures, the screened
deformation potential gives rise to $T^4$-term and the unscreened
deformation potential leads to $T^2 \ln T$-term. At low
temperatures the interference of electron-electron and
electron-impurity scattering dominates in the
temperature-dependent conductivity. At higher temperatures the
conductivity is determined by the electron-phonon-impurity
interference, which prevails over pure electron-phonon scattering
in a wide temperature range, which extends with increasing
disorder.
\end{abstract}
\pacs{PACS   numbers: 72.10.D}

\maketitle

\section{INTRODUCTION}

Interference of electron scattering mechanisms changes drastically
transport properties of disordered conductors. It violates the
Mathiessen rule, according to which the contributions to
conductivity due to random potential and electron-electron
(electron-phonon) interactions are additive. Additional
interference terms in conductivity have various temperature
dependencies. In the diffusion limit, $T \tau << 1$ ($\tau$ is the
elastic mean free time), Altshuler-Aronov corrections to
conductivity have been intensively studied in three-dimensional
conductors ($T^{1/2}$-term) and in two-dimensional structures
($\ln T$-term). \cite{AA}

Interference of electron-electron (electron-phonon) scattering and
elastic scattering from random potential also modifies
significantly the electron transport in the quasi-ballistic limit:
$T\tau \gg 1$ for electron-electron interaction and $q_Tl \gg 1$
for electron-phonon interaction ($q_T$ is the momentum of a
thermal phonon, $l=v_F \tau$ is the electron-mean free path. In
weakly disordered conductors the interference corrections are
always proportional to the Drude conductivity.
Electron-phonon-impurity interference in metals was theoretically
studied in our paper. \cite{RS1} In the quasi-ballistic limit, $T>
u/l$ ($u$ is the sound velocity), the interference correction to
conductivity is quadratic in the electron temperature,
\begin{eqnarray}\label{eph}
{\delta_{eph}\sigma \over \sigma_3} =\biggl[1-{\pi^2\over 16} - 2
\biggl( {u_l \over u_t} \biggr)^3 \biggl] \ {2 \pi^2 \beta_l T^2
\over 3 \epsilon_F p_F u_l},
\end{eqnarray}
where $\sigma_n= e^2 v_F^2 \tau \nu_n /n$ is the Drude
conductivity in corresponding dimensionality $n$, $\epsilon_F$ and
$p_F$ are the Fermi energy and momentum, $u_l$ and $u_t$ are the
longitudinal and transverse sound velocities, $\beta_l$ is the
constant of the electron-phonon interaction (see Sec. IV), and
$\nu_n$ is the electron density of states. It is interesting that
the longitudinal phonons give rise to a positive correction to
conductivity, while transverse phonons result in a negative
correction, which dominates in the temperature-dependent
conductivity due to stronger coupling of transverse phonons. This
$T^2$-term proportional to the Drude conductivity have been
observed in a wide temperature range, from 20K up to ~200K, in Nb,
Al, Be \cite{Pt}, NbC \cite{Il}, NbN \cite{Se1}, and W \cite{W}
films.

Electron-phonon interaction via the piezoelectric potential in
weakly disordered heterostructures was considered in the paper.
\cite{R1} It has been found that at low temperatures, T$\leq$
0.5K, where the piezoelectric potential is strongly screened, the
interference correction is given by
\begin{eqnarray}\label{piezo}
{\delta_{pz}\sigma \over \sigma_2} \simeq - {(eh_{14})^2 \over
4\rho u^3} {T^2 \over (\kappa_2 v_F)^2},
\end{eqnarray}
where $h_{14}$ is piezoelectric constant, $\kappa_2^{-1}$ is the
screening length, and $\rho$ is the density.

Investigation of the electron-electron interaction in weakly
disordered 2D electron systems has a long story.
\cite{Gold,R1,Sarma,R2,Al} After series of improvements, all
exchange (Fock) and direct (Hartree) processes have been taken
into account in the frame of the Landau Fermi-liquid theory in the
paper. \cite{Al} In the quasiballistic limit, $T\tau
> 1$, the correction to conductivity is
\begin{eqnarray}\label{Al}
{\delta_{ee}\sigma \over \sigma_2} =  \ \biggl( 1 + {3F_0^\sigma
\over 1+ F_0^\sigma }\biggr) \ {T \over \epsilon_F },
\end{eqnarray}
where $F_0^\sigma$ is the Fermi-liquid parameter describing
interaction in the triplet channel. Results of recent measurements
in GaAs/GaAsAl  heterostructures \cite{Sav,Sarma2} and Si MOSFETs
\cite{Ger,Vit} have shown good agreement with the theory \cite{Al}
at subkelvin (GaAs/GaAlAs) and helium (Si) temperatures. At higher
temperatures, the electron-phonon interaction dominates over
electron-electron scattering. While electron-phonon processes are
also very sensitive to disorder, the deformation electron-phonon
interaction in quasi-ballistic regime has not been studied to
date.

We would like to stress, that the interference corrections due to
the electron-phonon or electron-electron interactions always
originate from the {\it elastic part} of the corresponding
collision integral.\cite{AA, RS1, Al} Therefore, the interference
corrections depend on the electron temperature only. Early
theoretical papers on the electron-phonon-impurity interference
considered inelastic scattering from vibrating impurities and
extracted the $T^2$-correction to conductivity from the inelastic
part of the collision integral. \cite{T2} However, as it is shown
in our previous work, \cite{RS1} such terms cancel out and this is
a reason why the $T^2$-term is independent on the phonon
temperature.

In the current paper we continue studying weakly disordered
conductors. We calculate various interference corrections not
considered before. In Sec. II we start with the basic equations
describing interference phenomena in the electron transport. In
Sec. III we calculate the electron-electron corrections to the
conductivity in various dimensions with respect to the effective
interaction. The cross-over to the lower dimensionality occurs
when one of the conductor dimensions becomes smaller than
$q_c^{-1}$, where $q_c$ is the characteristic value of the
transferred electron momentum. For the electron-electron
interaction in weakly disordered conductors, $q_c^{-1}$ is of the
order of $L_T=v_F/T$. At sub-Kelvin and helium temperatures, $L_T
\sim 1-10 \mu m$, and the transition to the quasi-two-dimensional
case happens in relatively thick films with the three-dimensional
electron spectra and electron screening. The transition to the
quasi-one-dimensional case occurs in wires of radius $r \sim L_T$
and in 2D conducting channels of width $b \sim L_T$. We will show
that the interference corrections to the conductivity is mainly
determined by the sample dimensionality with respect to the
effective interaction. Dimensionality of the electron spectrum
just slightly changes numerical coefficients of the interference
corrections.

In Sec. IV we calculate corrections to conductivity due to the
electron-phonon interaction. We study interaction of electrons by
means of virtual phonons and interaction of electrons with thermal
phonons in bulk semiconductors and low-dimensional structures.
Considering the electron-phonon scattering, we will assume good
matching between the conducting and buffer layers and limit our
consideration to three-dimensional phonons. In Conclusions, we
summarize our main results, compare different terms, and discuss
experimental identification of interference contributions to
conductivity.

\section{ BASIC EQUATIONS}

Effects of interference between scattering mechanisms on the
electron transport can be studied by the linear response method as
well as by the quantum transport equation. Both methods are based
on the digrammatic technique.  The linear response method requires
many diagrams to be considered, while the transport equation deals
only with the electron self-energy diagrams but includes specific
terms in the form of Poisson brackets. \cite{RS1,R1}

In this paper we investigate the interference electron processes,
which are characterized by the momentum transfer much smaller than
the Fermi momentum. These processes can be described in the frame
of the Landau Fermi-liquid theory. The corresponding self-energy
diagrams for weakly disordered systems are shown in Fig. 1 and the
diagrams of the linear response method are presented in Fig. 2.
Results of the papers, \cite{RS1,R1,Al} which are briefly
reproduced in Appendix, show that in the quasiballistic limit the
correction to conductivity may be presented as
\begin{eqnarray}\label{sigma}
{\delta_{int}\sigma^{(k)} \over \sigma_n} = 2 \int {d\omega \over
2\pi} { d^k q \over (2\pi)^k} \ {f (\omega) \over \omega^2} \ \Im
 \Upsilon_n^{(k)}(q,\omega),
\end{eqnarray}
where $k$ is the dimensionality of the conductor with respect to
the interaction and $n$ is used for the dimensionality of the
electron spectrum. As we discussed in Introduction, the
dimensionality with respect to the effective electron-electron
interaction is determined by the characteristic length
$L_T=v_F/T$.

The function $f(\omega)$ is given by
\begin{eqnarray}\label{f}
 f(\omega) = {\partial \over \partial \omega} \biggl[\omega \coth \biggl(
  {\omega \over 2T} \biggr) \biggr].
\end{eqnarray}

The function $\Upsilon (q, \omega)$ is given by
\begin{eqnarray}\label{eF}
\Upsilon_n^{(k)}(q, \omega) = (\omega \tau)^2 \ V_n^R({\bf q},
\omega) \Phi_n^{(k)}(q, \omega),
\end{eqnarray}
where $V_n^R({\bf q}, \omega)$ is the retarded propagator
describing electron-electron or electron-phonon interaction, and
 $\Phi(q, \omega)$ is given by
\begin{eqnarray}\label{phi'}
 &&\Phi_n^{(k)}(q, \omega) = -{n \over v_F^2 \ \tau^2} \  \biggl[ \ \biggl\langle \biggl\langle
{\gamma^2 ({\bf v}_F {\bf e})^2 \over ({\bf q}{\bf v}_F -\omega
-i0)^2}\biggr\rangle_{\bf v} \biggl\rangle_{\bf q} \nonumber
\\ &&- \biggl\langle \biggl\langle {\gamma ({\bf v}_F {\bf e})^2
\over {\bf q}{\bf v}_F  -\omega  -i/\tau}\biggr\rangle_{\bf v}
\biggr\rangle_{\bf q} \cdot \biggl\langle {\gamma \over {\bf q}
{\bf v}_F  -\omega  -i0}\biggr\rangle_{\bf v} \nonumber
\\&&+ \biggl\langle \biggl\langle \ {\gamma \ {\bf v}_F
{\bf e} \over {\bf q}{\bf v}_F  -\omega -i0} \biggr\rangle_{\bf
v}^2 \biggr\rangle_{\bf q} \ \biggr]. \ \ \ \ \ \
\end{eqnarray}
In Eq. \ref{phi'}, $\gamma$ is the vertex of the electron-electron
or electron-phonons interaction, ${\bf e} = {\bf E}/E$ is the unit
vector in the direction of the electric field, and $\langle
\rangle_{{\bf v}({\bf q})}$ stands for the averaging over the
directions of ${\bf v}_F$ and ${\bf q}$.

For isotropic conductors (n=k), the averaging over the ${\bf p}$
and ${\bf q}$ directions is reduced to the averaging over the
angle $\phi$, the angle between ${\bf p}$ and ${\bf q}$. In this
case, Eq. \ref{phi'} can be simplified,
\begin{eqnarray}\label{phi}
&& \tau^2 \Phi_n^{(n)}(q, \omega)  = - \biggl\langle{\gamma^2
\over ({\bf q}{\bf v}_F -\omega  -i0)^2}\biggr\rangle_\phi
\nonumber \\ && + \biggl\langle {\gamma \over {\bf q}{\bf v}_F
-\omega -i0}\biggr\rangle_\phi ^2  -  \biggl\langle \ {{\bf q}{\bf
v}_F\over qv_F} \ {\gamma  \over {\bf q}{\bf v}_F -\omega -i0}
\biggr\rangle_\phi ^2 . \ \ \ \
\end{eqnarray}
Note, that in this form Eq. \ref{phi} is applicable to any
dimensionality of the electron system,
\begin{eqnarray}\label{angle}
\langle \psi ({\bf q}{\bf v}_F) \rangle_\phi =
\cases{{\displaystyle \int_{-1}^1 \psi ( q v_F x) \ {dx \over 2}}
, & \ 3D; \cr {\displaystyle \int_0^{2\pi} \psi ( q v_F \cos \phi
) \ { d\phi\over 2\pi}} & \ 2D; \cr {\displaystyle {1\over 2} \
\sum_{x=\pm 1} \psi ( q v_F x)} & \ 1D, \cr }
\end{eqnarray}
where $x = \cos (\phi)$.

In the next sections we use equation derived in this section to
calculate the quantum corrections to conductivity due to
electron-electron and electron-phonon interactions in weakly
disordered conductors of various dimensionality with respect to
the electron spectrum ($n$) and with respect to the interaction
($k$).

\section{ ELECTRON-ELECTRON INTERACTION}

As we discussed in the introduction, the effective
electron-electron interaction in weakly disordered conductors is
characterized by the momentum transfer of the order of $T/v_F$,
which is much smaller than the Fermi momentum (see also
calculations below). Therefore, the electron transport can be
described in the frame of the  Landau Fermi-liquid
theory.\cite{Al} In the singlet channel the bare interaction is
given the sum of the Coulomb potential,
\begin{eqnarray}\label{bare}
V_0(q) = \cases{ {\displaystyle {4\pi e^2 \over q^2} }, & \ 3D;
\cr {\displaystyle {2 \pi e^2 \over |q|} }, & \ 2D; \cr
{\displaystyle e^2 \ln {1 \over q^2 r^2}}, & \ 1D, \cr}
\end{eqnarray}
and the Fermi-liquid interaction,
\begin{eqnarray}\label{FL}
V_F \approx - F_0^{p}/\nu_n.
\end{eqnarray}

The screened interaction in the random phase approximation, which
is justified for small momentum transfers, is given by
\begin{eqnarray}\label{V}
V_n^{R(A)}(q,\omega) = {\nu_n V_0(q) - F_0^{p}  \over \nu_n-
[\nu_n V_0(q)-F_0^{p}] \ P_n^{R(A)}(q,\omega)}, \ \ \ \
\end{eqnarray}
where $P^{R(A)}(q,\omega)$ is the polarization operator.

In the absence of the magnetic field and spin-orbit scattering the
screened propagator in the triplet-channel may be taken in the
form \cite{Al}
\begin{eqnarray}\label{TrV}
 V_n^{R(A)}({\bf q},\omega) = -{ 3F_0^\sigma \over \nu_n - F_0^\sigma P_n^{R(A)}
 ({\bf q},\omega)},
\end{eqnarray}
where $F_0^\sigma$ is the Fermi-liquid constant. The above
equation assumes that the Fermi-liquid coupling is independent on
electron momenta. Restrictions of this approximation were
discussed in the paper\cite{Al}.

In the Subsecs. A and B we calculate the conductivity corrections
in the singlet channel, the triplet-channel corrections will be
studied in Subsec. C.

\subsection{Systems with three-dimensional spectrum}

First we consider a conductor with three-dimensional electron
spectrum. For $1/\tau \ll \omega \lesssim qv_F \ll \epsilon_F$,
the polarization operator is given by
\begin{eqnarray}\label{P}
P_3^R(q,\omega) = -\nu_3 \ \biggl[1 - {\displaystyle {\omega \over
qv_F} \ {\rm arctanh}\biggl({qv_F\over \omega+i0 }
\biggr)}\biggr],
\end{eqnarray}
where the branch of ${\rm arctanh}(y)$ is chosen as
\begin{eqnarray}\label{ar}
 {\rm arctanh}(y)= -{\pi i \over 2} +{1\over 2} \ \ln {y+1 \over y-1}, \ \  y>1.
\end{eqnarray}
Thus, the screened Coulomb potential may be presented as
\begin{eqnarray}\label{V1}
V_3^R(q,\omega)= \ { {\displaystyle {1 \over \nu_3} \ \biggr(
1-F_0^p {q^2\over \kappa_3^2} \biggr)} \over {\displaystyle
{q^2\over \kappa_3^2} +
 \biggr( 1-F_0^p {q^2\over \kappa_3^2} \biggr)
\biggl[1 -  {{\rm arctanh}(qv_F/\omega )\over
qv_F/\omega}}\biggr]}\nonumber \\ \ \ \ \ \,
\end{eqnarray}
where $\kappa_3^2 = 4\pi e^2 \nu_3$, and $\nu_3 = mp_F/ \pi^2$. In
the limit of strong screening, $\kappa_3 \gg q$, the screened
potential is independent on the form of the bare potential (the
unitary limit).

Taking into account that for the electron-electron interaction
$\gamma=1$ and calculating integrals in Eq. \ref{phi}, we find
\begin{eqnarray}\label{eephi}
\Phi_3^{(3)}(q, \omega) &=& {1\over \tau^2} \biggl[ {1\over
(qv_F)^2 -(\omega)^2}+ \biggl({{\rm arctanh}(qv_F/\omega )\over
qv_F}\biggr)^2 \nonumber \\ &-& {1\over (qv_F)^2} \biggl(1 - {{\rm
arctanh}(qv_F/\omega )\over qv_F/\omega  }\biggr)^2\biggl].
\end{eqnarray}
Substituting $\Phi(q, \omega)$ and $V^R(q, \omega)$ into Eq.
\ref{sigma}, we get the correction to the conductivity in the
singlet channel. For simplicity we present further results in the
limit $\kappa_3 \gg q$. Then,
\begin{eqnarray}\label{ee3d}
{\delta_{ee}^s \sigma \over \sigma_3} = 2 \int {d\omega \over
2\pi} \ {f(\omega)\over \omega^2} \int_{|\omega|/v_F}^{2p_F} {q^2
\ dq \over 2 \pi^2} \ \Im \Upsilon_3^{(3)}\biggl( {qv \over \omega
} \biggr),
 \ \ \ \ \ \
\end{eqnarray}
where
\begin{eqnarray}\label{F3}
&& \Upsilon_3^{(3)}(y) ={1\over \nu_3} \ {1 \over 1 - {\rm
arctanh}(y) / y} \nonumber
\\ &&\times \ \biggl[{1 \over y^2 - 1}+\biggl({{\rm arctanh} \ y
\over y} \biggr)^2 -\biggl({1\over y}-{{\rm arctanh} \ y\over
y^2}\biggr)^2\biggr]. \nonumber \\
\end{eqnarray}
Note, the low limit in the integral in Eq. \ref{ee3d} is chosen
taking into account that the imaginary part of $\Upsilon(qv/\omega
)$ exists only at $y=qv_F/\omega \geq 1$.

The function $\Upsilon_3 (qv_F/\omega)$ has the following
asymptotes
\begin{eqnarray}\label{Asym}
\Im \Upsilon_3^{(3)}(y) = {1 \over \nu_3} \cases{ \displaystyle{
\biggl({\pi^3\over 8} - 2\pi\biggr){1\over y^3}}, \ \ \ \ \ \ \ \
\ \ \ y \rightarrow \infty , \cr \displaystyle{ - {\pi \over
(\ln(y-1))^2} \ {1\over (y-1)}}, \ \  y \rightarrow 1 . \cr}
\end{eqnarray}
Therefore, with logarithmic accuracy we get
\begin{eqnarray}\label{F3int}
{v_F^3 \nu_3 \over \omega^3} \int_{|\omega|/ v_F}^{2p_F} dq \ q^2
\Im \Upsilon_3^{(3)} \biggl( {qv_F \over \omega } \biggr)  =
\biggl({\pi^3\over 8} - 2\pi\biggr)  \ln {4 \epsilon_F \over \mid
\omega \mid}. \nonumber \\
\end{eqnarray}
As seen from Eqs. \ref{F3int}, the integral over $q$ covers the
interval from $\omega/v_F \sim T/v_F$ to $2p_F$, but the
temperature dependence arises only from the low limit, $q=T/v_F$.
Therefore, the approximation, $q<<p_F$, which we made for the
polarization operator (Eq. \ref{P}) is justified by the
logarithmic accuracy of the integral in Eq. \ref{F3int}.

Finally, taking into account that
\begin{eqnarray}\label{poomega}
\int d\omega \ \omega f(\omega) \ \ln {\epsilon_F \over \mid
\omega \mid} = -{2 \pi^2 T^2 \over 3} \ln {4 \epsilon_F \over T},
\end{eqnarray}
we find
\begin{eqnarray}\label{final1}
{\delta_{ee}^s\sigma^{(3)}\over \sigma_3} &=& C_3 \biggl({T \over
\epsilon_F } \biggr)^2 \ln {4 \epsilon_F \over T}, \\ C_3 &=&
{\pi^2 \over 6} \ \biggl(1- {\pi^2 \over 16} \biggr).
\end{eqnarray}

We would like to remind that this result has been obtained in the
limit of strong screening, $\kappa_3 \gg 2p_F$, where the screened
potential does not depend on the bare potential. Our calculations
show that in the general case,  $\kappa_3 \sim 2p_F$, the leading
term, $T^2 \ln T$, is also given by Eq. \ref{final1}, while
additional terms are proportional to $T^2$.

Next we consider a quasi-two-dimensional conductor with 3D
electron spectrum. For a film or conducting layer of finite
thickness $d$, integration over the transverse component of the
wavevector $q$ in Eq. \ref{sigma} is replaced by summation over
eigenstates in the film. As seen from Eqs. \ref{F3int} and
\ref{poomega}, the characteristic value of $\omega$ is of the
order of $T$ and the characteristic values of the momentum $q$ is
$\omega/v_F$. Therefore, a transition to quasi-two-dimensional
case occurs, if the thickness of a conducting layer , $d$, is of
the order of the characteristic length $L_T= v_F/T$. For a
quasi-two-dimensional conductor, $d<<L_T$, the electron
transitions with $q_\perp = 0$ should be retained,
\begin{eqnarray}\label{sum}
{d^3 q\over (2\pi)^3} \longrightarrow {1\over d} {d^2 q \over
(2\pi)^2} \ \delta_{q_\perp , 0}.
\end{eqnarray}
Averaging Eq. \ref{phi'} over the directions of vectors ${\bf q}$
and ${\bf v}_F$, one should take into account that vectors ${\bf
q}$ and ${\bf e}$ lie in the plane of the conductor, while the
vector ${\bf v}_F$ has an arbitrary direction. In this geometry we
get
\begin{eqnarray}\label{phi'32}
 \Phi_3^{(2)}(q, \omega) &=& -{3 \over \tau^2} \  \biggl[ \ {1\over 4}\biggl\langle
{ (1+x^2) \over (q v_F x -\omega -i0)^2}\biggr\rangle_{\phi}
\nonumber
\\ &-& {1\over 4} \biggl\langle {(1+x^2)
\over q v_F x  -\omega  -i/\tau}\biggr\rangle_{\phi} \cdot
\biggl\langle {1 \over q v_F x -\omega -i0}\biggr\rangle_{\phi}
\nonumber \\ &+& {1\over 2 }\biggl\langle \ {  x \over q v_F x
-\omega -i0} \biggr\rangle_{\phi}^2 \ \biggr],
\end{eqnarray}
where  $x=\cos \phi$ and the averaging over the angle $\phi$ is
given by Eq. \ref{angle}.

Calculating  $\Phi_3^{(2)}(q, \omega)$, in the limit of strong
screening we get the function $\Upsilon(qv_F/\omega)$ defined by
Eq. \ref{eF},
\begin{eqnarray}\label{F32}
&& \nu_3 \Upsilon_3^{(2)}(y) = {3\over 4 y^2 }\biggl(1 - {{\rm
arctanh} \ y \over y} \biggr)^{-1} \nonumber
\\ && \times \biggl[{4 - 2y^2\over y^2-1} + {5 {\rm arctanh} \ y \over
y} + (y^2-1)\biggl({{\rm arctanh}\ y \over y} \biggl)^2\biggr].
\nonumber \\
\end{eqnarray}

Thus, the correction to the conductivity (Eq. \ref{sigma}) is
given by
\begin{eqnarray}\label{quasi2D}
{\delta_{ee}^s \sigma^{(2)}\over \sigma_3} = -{1 \over \pi^2 d
v_F^2 \nu_3}\ C_2 \int_0^{\epsilon_F} d \omega   f(\omega),
\end{eqnarray}
where
\begin{eqnarray}\label{q2D}
C_2 = -\int_1^{\infty} dy \ y \ \nu_3 \ \Im \Upsilon_3^{(2)}(y)
\approx 5.3
\end{eqnarray}
Here the integral over $q$ can be extended to infinity, because
only processes with small momentum transfer, $q \sim \omega/ v_F$,
are important. Taking into account that
\begin{eqnarray}\label{poomega2}
\int_0^{\epsilon_F} d\omega \  f(\omega) = - 2T+\epsilon_F \coth
\biggl({\epsilon_F\over 2T}\biggr),
\end{eqnarray}
we see that the temperature dependent correction to the
conductivity is determined by the first term in Eq.
\ref{poomega2}. From Eq. \ref{quasi2D} we find
\begin{eqnarray}\label{final2}
{\delta_{ee}^s \sigma^{(2)} \over \sigma_3} =  \ {C_2 \over p_F d}
\ {T \over \epsilon_F }.
\end{eqnarray}
In Eq. \ref{final2} we omit additional term proportional to
$(T/\epsilon_F)^2 \ln(p_Fd)$. It originates from the interval $1/d
\ll q \leq 2p_F$ and it is small compared with the linear term.

Now we consider quasi-one-dimensional conductors,  such as wires
with radius $r$, which is much smaller than $L_T$. In this case
the integration over the transfer momentum in Eq. \ref{sigma} is
replaced by
\begin{eqnarray}\label{sum1}
{d^3 q\over (2\pi)^3} \longrightarrow {1\over \pi r^2} {d q \over
2\pi} \ \delta_{q_\perp , 0}.
\end{eqnarray}

In the quasi-one-dimensional case the vectors ${\bf q}$ and ${\bf
e}$ are parallel. Averaging Eq. \ref{phi'} over the angles of
${\bf q}$ we get
\begin{eqnarray}\label{phi'31}
 \Phi_3^{(1)}(q, \omega) = -{3 \over \tau^2} \  \biggl[ \ \biggl\langle
{ x^2 \over (q v_F x -\omega -i0)^2 } \biggr\rangle_{\phi}
\nonumber
\\ - \biggl\langle {x^2
\over q v_F x  -\omega  -i/\tau} \biggr\rangle_{\phi} \cdot
\biggl\langle {1 \over q v_F x -\omega -i0}
\biggr\rangle_{\phi}\biggr].
\end{eqnarray}
In the limit of strong screening, the function
$\Upsilon(qv_F/\omega)$ (Eq. \ref{eF}), is given by
\begin{eqnarray}\label{F31}
&& \nu_3 \Upsilon_3^{(1)}(y) = {3\over y^2} \biggl(1 - {{\rm
arctanh} \ y \over y} \biggr)^{-1}\nonumber
\\ && \times \biggl[{2 - 2y^2\over y^2-1} + {{\rm arctanh} \ y \over
y} + \biggl({{\rm arctanh}\ y \over y} \biggl)^2\biggr]. \ \ \ \ \
\end{eqnarray}
Then, the correction to the conductivity (Eq. \ref{sigma}) is
\begin{eqnarray}\label{quasi1D}
{\delta_{ee}^s \sigma^{(1)}\over \sigma_3} = -{1 \over \pi^3 r^2
v_F \nu_3}\ C_1 \int_0^{\epsilon_F} d \omega \ {f(\omega)\over
\omega},
\end{eqnarray}
where
\begin{eqnarray}\label{q1D}
C_1 = -\int_1^{\infty} dy \ y \ \nu_3 \ \Im \Upsilon_3^{(1)}(y)
\approx 4.3
\end{eqnarray}
Taking into account that
\begin{eqnarray}\label{poomega1}
\int_0^{\epsilon_F} d\omega \  {f(\omega)\over \omega} =
\ln\biggl({\epsilon_F\over 2T}\biggl),
\end{eqnarray}
finally we get
\begin{eqnarray}\label{final1D}
{\delta_{ee}^s \sigma^{(1)} \over \sigma_3} =\ {C_1 \over \pi (r
p_F)^2} \ \ln \biggl({2T \over \epsilon_F} \biggr).
\end{eqnarray}

\subsection{Two-dimensional electron spectrum}
Next we calculate the interference correction in conductors with
two-dimensional electron spectra. For 2D electron gas the
polarization operator in the quasi-ballistic limit is
\begin{eqnarray}\label{P2}
P^R_2(q,\omega) = -\nu_2 \ \biggl(1 - {\omega \over \sqrt{
(\omega+i0)^2-(q v_F)^2} }\biggr), \ \ \ \
\end{eqnarray}
where $\nu_2 = m/\pi$. Using Eq. \ref{phi}, in the limit of strong
screening, $V^R(q, \omega) = -1/P^R_2(q,\omega)$, we get\cite{Al}
\begin{eqnarray}\label{F2D}
\nu_2 \Upsilon_2^{(2)}(y) = {1\over 1-(y-i0)^2} + { 1 \over y^2
\sqrt{1 - (y-i0)^2} }, \ \ \ \ \ \ \
\end{eqnarray}
and the corresponding correction to the conductivity (Eq.
\ref{sigma}) is given by
\begin{eqnarray}\label{2D}
{\delta_{ee}^s \sigma^{(2)} \over \sigma_2} &=&  - {1\over \pi
v_F^2 \nu_2} \int_0^{\epsilon_F} d \omega  \ f(\omega).
\end{eqnarray}
The integral over $\omega$ in Eq. \ref{2D} is exactly the same as
in the case of quasi-two-dimensional conductor (Eq.
\ref{poomega2}). Finally, we get the correction to conductivity
given by Eq. \ref{Al}. As seen, in the quasi-ballistic limit the
polarization operator (Eqs. \ref{P} and \ref{P2}) and the function
$F(y)$ (Eqs. \ref{F32} and \ref{F2D}) has different forms for
quasi-two-dimensional and two-dimensional conductors. Therefore,
the final results (Eqs. \ref{Al} and \ref{final2}) differ by a
numerical factor.

Next we consider conductivity in the quasi-one-dimensional
conductor, such as a narrow channel with width $b<L_T$. Taking
into account that in the quasi-one-dimensional case the vectors
${\bf q}$ and ${\bf E}$ are parallel and averaging  Eq. \ref{phi'}
over the angles of ${\bf q}$, we get
\begin{eqnarray}\label{phi'21}
&& \Phi_2^{(1)}(q, \omega) = -{2 \over \tau^2} \  \biggl[ \
\biggl\langle { (\cos \phi)^2 \over (q v_F \cos \phi -\omega
-i0)^2 } \biggr\rangle_{\phi} \nonumber
\\ && - \biggl\langle {(\cos \phi)^2
\over q v_F \cos \phi  -\omega  -i/\tau} \biggr\rangle_{\phi}
\cdot \biggl\langle {1 \over q v_F \cos \phi -\omega -i0}
\biggr\rangle_{\phi}\biggr]. \nonumber \\
\end{eqnarray}
After averaging over the angle $\phi$, we find
\begin{eqnarray}\label{phi'21'}
 \Phi_2^{(1)}(q, \omega) &=& {2\over \tau^2} {1\over (\omega+i0)^2
-(qv)^2 }\nonumber \\ &\times& \biggr[ 1- {\omega \over
\sqrt{(\omega+i0)^2 -(qv)^2}}\biggl].
\end{eqnarray}

Therefore, in the unitary limit, the function
$\Upsilon(qv_F/\omega)$ (Eq. \ref{eF}), is given by
\begin{eqnarray}\label{F2D1}
\nu_2 \Upsilon_2^{(1)}(y) = {2\over 1-(y-i0)^2}.
\end{eqnarray}
Calculating the correction to the conductivity in the singlet
channel,
\begin{eqnarray}\label{2D1}
{\delta_{ee}^s \sigma^{(1)}\over \sigma_2} =  {1\over \pi^2 \nu_2
v_F b} \int_0^{\epsilon_F} d \omega \ {f(\omega) \over \omega}
\int_0^{\infty} dy \ \Im \Upsilon_2^{(1)}(y), \ \ \ \ \
\end{eqnarray}
we get
\begin{eqnarray}\label{q1D2}
{\delta_{ee}^s \sigma^{(1)} \over \sigma_2} \ = \ {1 \over  p_F  b
} \ \ln \biggl({2T\over \epsilon_F}\biggr).
\end{eqnarray}
Thus, in quasi-one-dimensional conductors with 3D electron spectra
(Eq. \ref{final1D}) and 2D spectra (Eq. \ref{q1D2}) the
corrections have the logarithmic temperature dependence.

\begin{table}
\caption{Corrections to the conductivity $\delta_{ee}^s
\sigma^k/\sigma_n$ due to electron-electron interaction in the
singlet channel.}
\begin{ruledtabular}
\begin{tabular}{cccc}
  n / k
\footnote{$n$ is the dimensionality of a conductor with respect to
electron spectrum, $k$ is the dimensionality with respect to the
interaction (crossover to low dimensionality occurs at the
characteristic length $L_T = v_F/T$); $\sigma_n$ is the Drude
conductivity; $d$ is the thickness of a quasi-two-dimensional
conductor, $r$ is the radius of a quasi-one-dimensional conductor,
$b$ is the width of a conductor curved from 2D-layer; $C_1, C_2,$
and $C_3$ are numerical constants defined by Eqs. \ref{q1D},
\ref{q2D}, and \ref{final1}.}
  & k=3 & k=2 & k=1 \\[6pt] \hline \\[6pt]
  n=3 &   \ ${\displaystyle {C_3 T^2 \over \epsilon_F^2 }
  \ln {4 \epsilon_F \over
T}}$ \ & \ ${\displaystyle {C_2 \over p_F d} \ {T \over \epsilon_F
}}  $ & \ ${\displaystyle {C_1 \over \pi (r p_F)^2} \ \ln {2T
\over \epsilon_F}}$
\\[15pt]
  n=2 & - &  \ ${\displaystyle {T \over \epsilon_F }}$ &
\ ${\displaystyle {1 \over  p_F  b } \ \ln {2T\over \epsilon_F}}$
\\[15pt]
\end{tabular}
\end{ruledtabular}
\end{table}

\vskip 12pt The main results of Subsecs. A and B are summarized in
Tab. 1. In these subsections we have considered the
singlet-channel interaction in the limit of strong screening (the
unitary limit), which gives the upper bound for the strength of
the interelectron interaction. As it follows from calculations
above, the characteristic value of the electron momentum
transferred is of the order of $T/v_F$. Thus, the leading
corrections to conductivity are accumulated at large distances,
$L_T \simeq \hbar v_F /k_B T$, where the electron-electron
interaction is strongly screened. In the unitary limit the
interaction has an universal form, which is independent of the
original interaction and its renormalization by the Fermi liquid
parameters.

\subsection{Triplet channel interaction}

Conductivity corrections in the triplet channel are calculated in
the same way as the singlet-channel corrections.  Substituting the
triplet-channel propagator (Eq. \ref{TrV}) and the function
$\Phi_3^{(3)}$ (Eq. \ref{eephi} in Eq. \ref{eF}) in to Eq.
\ref{eF}, we find the function  $\Upsilon(y)$ for a bulk
conductor,
\begin{eqnarray}\label{TrF3}
 && \Upsilon_3^{(3)}(y) ={1\over \nu_3} \ {3 F_0^\sigma \over 1 +
F_0^\sigma \biggl(1 - {\rm arctanh}(y) / y \biggr)} \nonumber
\\ && \times \ \biggl[{1 \over y^2 - 1}+\biggl({{\rm arctanh} \ y
\over y} \biggr)^2 -\biggl({1\over y}-{{\rm arctanh} \ y\over
y^2}\biggr)^2\biggr]. \nonumber \\
\end{eqnarray}
Finally, integrating Eq. \ref{ee3d}, with logarithmic accuracy we
get
\begin{eqnarray}\label{Tr3}
{\delta_{ee}^t \sigma^{(3)}\over \sigma_3} &=& {3F_0^\sigma \over
1+F_0^\sigma} \biggl(1- {\pi^2\over 16} \ {F_0^\sigma \over
1+F_0^\sigma} \biggr) \nonumber \\ &\times& {\pi^2\over 6}
\biggl({T \over \epsilon_F } \biggr)^2 \ln {4 \epsilon_F \over T}.
\end{eqnarray}
Results for the quasi-two and quasi-one-dimensional conductors
with 3D spectrum cannot be presented in analytical form and will
not be considered here.

In the case of the quasi-one-dimensional conductor carved from the
two-dimensional structure (see Subsec. B), the function
$\Upsilon(y)$ is given by
\begin{eqnarray}\label{TrF1}
\Upsilon_2^{(1)}(y) &=& - {1\over \nu_2} \ {3F_0^\sigma \over
1+F_0^\sigma} \ {2\over 1-(y-i0)^2} \nonumber \\ &\times&
\biggl(1-{1\over (1+F_0^\sigma) \sqrt{1-y^2} -F_0^\sigma} \biggr).
\end{eqnarray}
Finally, integrating Eq. \ref{2D1}, we get
\begin{eqnarray}\label{Tr1}
{\delta_{ee}^t \sigma^{(1)} \over \sigma_2} \ = {3
\biggr(F_0^\sigma + \pi  G(F_0^\sigma)\biggl) \over 1+F_0^\sigma}
\ {1\over p_F b } \ \ln \biggl({2T\over \epsilon_F}\biggr),
\nonumber \\
\end{eqnarray}
where
\begin{eqnarray}\label{cases}
&&  G(F_0^\sigma) = -1 - {\displaystyle {2\over \pi} {\mid
1+F_0^\sigma \mid \over \sqrt {\mid 1+2F_0^\sigma \mid }}}
\nonumber
\\ && \times \cases{ {\displaystyle {\rm arctanh}
{\sqrt{-1-2F_0^\sigma } \over F_0^\sigma } }, & $F_0^\sigma <
-{1\over 2}$, \cr {\displaystyle \arctan{{\sqrt{1+2F_0^\sigma }
\over  F_0^\sigma}}}, &  $-{1\over 2} < F_0^\sigma < 0$, \cr
{\displaystyle \arctan{ {\sqrt{1+2F_0^\sigma} \over  F_0^\sigma} }
-\pi}, &  $F_0^\sigma > 0 $. \cr} \ \ \ \ \
\end{eqnarray}
Thus, the temperature dependence of the conductivity corrections
in the triplet channel (Eqs. \ref{Tr3} and \ref{Tr1}) is the same
as in the singlet channel, but the value and sign depend on the
parameter $F_0^\sigma$.

\section{ELECTRON-PHONON INTERACTION}
To apply Eqs.
\ref{sigma} and \ref{phi} to the electron-phonon interaction one
should specify the phonon propagator and electron phonon vertex.
The retarded phonon Green function is given by
\begin{eqnarray}\label{Ph}
 D^R({\bf q},\omega) =
(\omega-\omega_{\bf q} + i0)^{-1} - (\omega+\omega_{\bf q} +
i0)^{-1}.
\end{eqnarray}

The unscreened vertex of the electron-phonon scattering due to the
deformation potential is
\begin{eqnarray}\label{D}
 \gamma = \ { D {\bf q}\cdot{\bf e}_n  \over (2 \rho
 \omega_q)^{1/2}},
\end{eqnarray}
\noindent where ${\bf e}_n$ is the phonon polarization vector.

In the isotropic model, for longitudinal phonons the deformation
potential is described by two constants $D_0$ and $G$, \cite{Gant}
\begin{eqnarray}\label{DG}
D=D_0-3 G(\cos \theta)^2.
\end{eqnarray}

Screening of the bare electron-phonon vertex presented in Fig. 3
leads to
\begin{eqnarray}\label{gammas}
\gamma_{sc} &=& \ {q  \over (2 \rho
 \omega_q)^{1/2}} \ \biggl[{D_0 \over 1-V_0(q)P^R(q,\omega)}
\nonumber
\\ &-& 3G\biggl((cos\theta)^2 + { V_0(q)P_2^R(q,\omega)\over
1-V_0(q)P(q,\omega)}\biggr)\biggr] ,
\end{eqnarray}
where $P(q,\omega)$ is given by Eq. \ref{P}, and $P_2(q,\omega)$
is the electron polarization operator with the vertex
$(\cos\theta)^2$.

In a semiconductor, the deformation potential constant $D_0$ is
much larger then the constant $G$, which has strong concentration
dependence (see below). For this reason, the first term in the
square brackets in Eq. \ref{gammas} is only taken into account.
For thermal phonons $qv_F$ is much larger than  $\omega$, and,
therefore, the dynamic part of $P(q,\omega)$ proportional to
$\omega$ may be neglected. Therefore, in a semiconductor
interaction between electrons and real phonons is described by the
vertex
\begin{eqnarray}\label{gammasem}
\gamma_{sem} = {q  \over (2 \rho
 \omega_q)^{1/2}} {D_0 \over 1+ \kappa_3^2/q^2}.
\end{eqnarray}

 In a
metal two constants of the deformation potential $D_0$ and $G$ are
of the same order. However, due to the strong screening
($V_0(q)P^R(q,\omega) >> 1$), the term with the constant $D_0$ in
Eq. \ref{gammas} becomes negligible \cite{Gant}. Thus, for a metal
we get
\begin{eqnarray}\label{gam}
\gamma_{met} &=& {G \ q  \over (2 \rho
 \omega_q)^{1/2}} \biggl[\biggl(1 - {\displaystyle {\omega \tau
\over ql} \ {\rm arctanh}\biggl({ql\over \omega \tau}
\biggr)}\biggr)^{-1} \nonumber \\ &+& 3\biggl({\omega \over qv}
\biggr)^2 -3(\cos\theta)^2 \biggr],
\end{eqnarray}
In the static limit, $ql >> \omega \tau$, which is applicable to
the thermal phonons, the last equation reduces to
\begin{eqnarray}\label{met}
\gamma_{met}= [1-(3 \cos \theta)^2] \ {G \ q  \over (2 \rho
 \omega_q)^{1/2}},
\end{eqnarray}
where $G=(2/3)\epsilon_F$ \cite{RS1,Gant}.

Vertices obtained in this section will be used to calculate
interference correction to conductivity due to virtual phonons
(Sec. V) and thermal phonons (Sec. VI).

\subsection{ Virtual phonons}

Besides the Coulomb potential, the electron-electron interaction
may be realized via intermediate electron-ion interaction, i.e.
via virtual phonons. If $\omega_q=qu \gg \omega \sim T$, according
to Eq. \ref{Ph}, the phonon propagator $D^R$ is real and equal to
$-2/\omega_q$. Then, from Eq. \ref{sigma} we get
\begin{eqnarray}\label{sigma1}
{\delta_{e-v.ph}\sigma \over \sigma_3} = -4 \tau^2 \int {d\omega
\over 2\pi} {q^2 dq \over 2\pi^2} \ f (\omega) \ \omega_q^{-1} \
\Im \Phi(q, \omega).
\end{eqnarray}
We start with the electron-phonon interaction due to the
deformation potential ($D_0$) in a bulk semiconductor. In this
case the square of the electron-phonon vertex does not have an
imaginary part, and the function $\Im  \Phi(q, \omega)$ is given
by Eq. \ref{eephi}. Integrating Eq. \ref{sigma1} over $q$, one
should take into account the characteristic interval $T/u < q <
2p_F$. Therefore, with logarithmic accuracy we get
\begin{eqnarray}\label{final4}
{\delta^{sem}_{e-v.ph} \sigma \over \sigma_3} = -{2 \over 3}{D_0^2
T^2\over \rho v_F^3 u^2} \ln \biggl({p_F u \over T} \biggr) .
\end{eqnarray}

In  metals the deformation potential is effectively screened, and
in $\Im  \Phi(q, \omega)$ one should take into account the
imaginary part of the electron-phonon vertex (Eq. \ref{gam}).
Calculating the integrals in Eq. \ref{phi} with the vertex
$\gamma_{met}$ (Eq. \ref{gam}) we find
\begin{eqnarray}
\Im \Phi(q, \omega)= -2\pi\biggl({\pi^2\over 8}-1 \biggr) \
{\omega \tau \over ql}.
\end{eqnarray}
Substituting this result into Eq. \ref{sigma1} we get
\begin{eqnarray}\label{vphmet}
{\delta^{met}_{e-v.ph}\sigma \over \sigma_3} = {8 \over
27}\biggl({\pi^2\over 8 }-1\biggr) \ {\epsilon_F^2 \ T^2 \over
\rho v_F^3 u^2}  \ln \biggl({p_Fu\over T} \biggr).
\end{eqnarray}
Note, that large electron momentum is transferred via exchange of
virtual phonons. According to Sec. III, such processes are
important only in conductors with 3D electron spectrum. In 2D
structures, the electron-electron correction is associated with
processes of small momentum transfer, therefore, the contribution
of virtual phonons is absent.

\subsection{Thermal Phonons}

For thermal phonons, the imaginary part of the phonon propagator
(Eq. \ref{Ph}) is only important,
\begin{eqnarray}\label{PhTh}
\Im D^R(q, \omega) = i\pi [\delta(\omega -\omega_q) -
\delta(\omega+\omega_q)].
\end{eqnarray}
Taking into account that $\omega$ is of the order of $T$ and much
smaller than $qv_F$, we can put $\omega=0$ in the function
$\Phi(q, \omega)$ (Eq. \ref{phi}). After integrating  over
$\omega$, Eq. \ref{sigma} takes the form
\begin{eqnarray}\label{sigma2}
{\delta_{e-ph} \sigma \over \sigma_3} = -2 \tau^2 \int {q^2 dq
\over 2\pi^2} \ f (\omega_q) \ \Re  \Phi(q),
\end{eqnarray}
where $\Phi(q)=\Phi(q, \omega=0)$.

In  metals, $\Re \Phi(q)$ is calculated with the electron-phonon
vertex $\gamma_{met}$ in the static limit (Eq. \ref{met}). From
Eq. \ref{phi}, we get
\begin{eqnarray}\label{phimet}
\Re \Phi(q) =  \biggl({16\over \pi^2}-1 \biggr)  \biggl({\pi \over
2ql} \biggr)^2 \biggl({2\over 3} \epsilon_F \biggr)^2 {q^2 \over 2
\rho \omega_q}.
\end{eqnarray}
Substituting this result into Eq. \ref{sigma2}, we find
\begin{eqnarray}\label{sigmamet}
{\delta_{e-ph}^{met}\sigma \over \sigma_3 } =
\biggl(1-{\pi^2\over 16} \biggr) \ {2 \pi^2 \beta_l T^2 \over
\epsilon_F p_F u_l},
\end{eqnarray}
where the electron-phonon interaction constant is
\begin{eqnarray}\label{beta}
\beta_l =  \biggl({2\over 3} \epsilon_F\biggr)^2 {\nu \over 2 \rho
u_l^2}.
\end{eqnarray}
 Note, that in
the isotropic model the deformation potential for transverse
phonons is $-(3/2)G \sin(2\theta)$. Taking into account the
interaction with transverse phonons, we will get Eq. \ref{eph}
obtained in the paper \cite{RS1}.

For a bulk semiconductor, Eq. \ref{phi} with the vertex
$\gamma_{sem}$ (Eq. \ref{gammasem}) results in
\begin{eqnarray}\label{phisem}
\Re \Phi(q) =  -  \biggl({\pi \over 2ql} \biggr)^2 \biggl({q^2
\over q^2+ \kappa_3^2}\biggr)^2 { D_0^2 q^2 \over 2 \rho
\omega_q}.
\end{eqnarray}
At low temperatures, $T < \kappa_3 u$, the deformation potential
is strongly screened. In this limit, the interference correction
to the conductivity is given by
\begin{eqnarray}\label{final5}
{\delta_{e-ph}^{sem}\sigma \over \sigma_3} = -{10 \pi^6 \over 63}{
D_0^2 T^6 \over \rho v_F^2 u^7 \kappa_3^4}.
\end{eqnarray}
At higher temperatures, $T > \kappa_3 u$, the deformation
potential is not screened. In this limit Eqs. \ref{sigma2} and
\ref{phisem} result in
\begin{eqnarray}\label{final6}
{\delta_{e-ph}^{sem} \sigma \over \sigma_3} = -{\pi^2 \over
24}{D_0^2 T^2 \over \rho v_F^2 u^3}.
\end{eqnarray}

For a two-dimensional electron gas in heterostructures interacting
with three-dimensional phonons, Eq. \ref{phi} takes the form
\begin{eqnarray}\label{phisem2}
\Re \Phi(q) =  -  {2 \over (q_\parallel l)^2} \biggl({q_\parallel
\over q_\parallel + \kappa_2}\biggr)^2 { D_0^2 q^2 \over 2 \rho
\omega_q},
\end{eqnarray}
where $q_\parallel$ is the wavevector component along the
conducting plane, $\kappa_{2}=2\pi e^2 \nu_2, \nu_2=m/\pi.$

Therefore, at low temperatures, $T < \kappa_2 u$, the correction
to conductivity is given by
\begin{eqnarray}\label{final7}
{\delta_{e-ph}^{sem}\sigma \over \sigma_2} = -{2 \pi^2 D_0^2 \over
5 \rho v_F^2 u^5 \kappa_2^2} \ T^4,
\end{eqnarray}
and, in the opposite limit,
\begin{eqnarray}\label{final8}
{\delta_{e-ph}^{sem} \sigma \over \sigma_2} = -{D_0^2 \over 3 \rho
v_F^2 u^3} \ T^2 \ln {T \over T_1},
\end{eqnarray}
where $T_1 = \max\{u \kappa_2, u / l \} $.

In semiconductors the constant $D_0$ is much larger than $G$,
therefore, in isotropic semiconductors interaction with
longitudinal phonons plays a key role. The interference correction
to conductivity is negative (Eqs. \ref{final5}, \ref{final6},
\ref{final7}, and \ref{final8}) and results in the resistivity
increases with temperature.

\begin{table}
\caption{Corrections to the conductivity
$\delta_{e-ph}\sigma/\sigma_n$ due to the interaction of electrons
with thermal phonons through the deformation potential.}
\begin{ruledtabular}
\begin{tabular}{cccc}
  \omit& D=3 & D=2  \\[6pt] \hline \\[6pt]
 $ T< \kappa_n u $  \ &   \ ${\displaystyle  -{10 \pi^6 \over 63} {
D_0^2 T^6 \over \rho v_F^2 u^7 \kappa_3^4} } $ \ & \
${\displaystyle -{2 \pi^2 D_0^2 \over 5 \rho v_F^2 u^5 \kappa_2^2}
\ T^4 }  $
\\[15pt]
 $ T > \kappa_n u $ & $ {\displaystyle  -{\pi^2 \over
24}{D_0^2 T^2 \over \rho v_F^2 u^3} } $ &  \ ${\displaystyle
-{D_0^2 \over 3 \rho v_F^2 u^3} \ T^2 \ln {T \over T_1} }$
\\[10pt]
\end{tabular}
\end{ruledtabular}
\vskip 2pt $\kappa_n$ is the inverse screening length, $D_0$ is
the deformation potential, $u$ is the sound velocity, and $\rho$
is the density.
\end{table}

Results obtained in this section have simple physical
interpretation. In the limit of weak disorder, $ql>1$, the
interference can be described by effective large-angle electron
scattering process with square of the matrix element
$a_{e-ph}^2/(ql)$, where $a_{e-ph}$ is the matrix element of the
'pure' electron-phonon scattering. Therefore, in the
Bloch-Gruneisen regime, the interference contribution to the
conductivity turns out to be $(p_F/q)^2/(ql)$ times larger than
the contribution of pure electron-phonon scattering. In the case
of the Debye phonon spectrum, the corresponding temperature
dependence of conductivity is modified by $T^{-3}$ factor.

As it follows from the interpretation above, the interference
corrections due to the electron-phonon interaction critically
depend on the dimension of phonons and phonon spectrum. For
example, flexural modes with dispersion $\omega_q \propto q^2$
dominate in the electron scattering in free standing thin films at
low temperatures. In the Bloch-Gruneisen regime, the conductivity
of pure films \cite{Glav} is proportional to $T^{7/2}$,  then in
weakly disordered films this dependence is proportional to $T^2$.

\section{CONCLUSIONS}

In this work we investigated  the interference of
electron-electron (electron-phonon) scattering and elastic
electron scattering from impurities and defects in weakly
disordered conductors and heterostructures. We have calculated the
interference corrections to the conductivity and demonstrated that
even weak disorder significantly modifies its temperature
dependence.

In weakly disordered conductors, characteristic momentum transfers
are of the order of $T/v_F$, which is significantly smaller than
the Fermi momentum.  Therefore, the Landau Fermi-liquid theory is
applicable and all processes with the large momentum transfer are
taken into account by the effective Fermi-liquid constants. Both
singlet and triplet channels of the electron-electron interaction
give interference corrections to conductivity,
\begin{eqnarray}\label{tot}
\delta_{ee} \sigma = \delta_{ee}^s \sigma + \delta_{ee}^t \sigma.
\end{eqnarray}
 Due to the Coulomb
potential divergence at small momenta, the singlet-channel
interaction corresponds to the unitary limit and corresponding
corrections are independent on the Fermi-liquid parameters (see
Tab. 1). The triplet-channel corrections (Eqs. \ref{Tr3} and
\ref{Tr1}) have the same temperature dependence as the
singlet-channel corrections. Contrary to the singlet channel, the
triplet channel corrections are not universal. Therefore, the
value and sign of the total correction depend on the Fermi-liquid
parameter $F_0^\sigma$ in the triplet channel. In the weak
coupling limit, $|F_0^\sigma| << 1$, the singlet-channel dominates
over the triplet one and the corrections to conductivity are
positive. Negative values of $F_0^\sigma$ may result in the
negative total correction, which is observed in
heterostructures.\cite{Ger}

Our main results for the singlet channel are summarized in Tab. I.
We found that in weakly disordered bulk conductors the
electron-electron interaction results in $T^2\ln T$ - term in
conductivity (Eq. \ref{final1}). In a quasi-two-dimensional
conductor, $d < v_F/T$, with 3D electron spectrum, $p_Fd \gg 1$,
the electron-electron interaction results in $T$-term (Eq.
\ref{final2}), which is the leading temperature-dependent term at
subkelvin temperatures. Our result differs from that for 2D
electrons \cite{Al} by a numerical factor. In the quasi-ballistic
case, integrals of electron Green functions for
quasi-two-dimensional conductors with 3D and 2D spectra (Eqs.
\ref{F32} and \ref{F2D1} ) are significantly different  and result
in different coefficients.

In  quasi-one-dimensional conductors with 3D and 2D spectra (wires
and channels), the interference corrections are proportional to
$\ln (T)$ (Eqs. \ref{final1D} and \ref{q1D2}). Note, that at
sub-Kelvin temperatures the characteristic length, $d_c = v_F/T $,
is of the order of 1 - 10 $\mu$m. Therefore, experiments with
wires and channels of $\mu$m-sizes would allow to observe
crossovers to lower dimensions. Note, that the logarithmic term
has been recently observed in arrays of open quantum dots of
$\mu$m-sizes at sub-Kelvin temperatures. \cite{Bird} This
observation may be relevant to the quasi-one-dimensional
interference corrections calculated in this paper.

We considered the electron-electron interaction via virtual
phonons and found that this interaction results in $T^2 \ln
T$-term (Eq. \ref{final4}). The interference corrections due to
interaction of electrons with thermal phonons are summarized in
Tab. II. In bulk semiconductors at low temperatures, $T < \kappa_3
u$, the contribution of thermal phonons interacting with electrons
via the screened deformation potential results in $T^6$-term (Eq.
\ref{final5}). At higher temperatures the interaction via
unscreened deformation potential results in $T^2$-term (Eq.
\ref{final6}). In two-dimensional heterostructures the screened
deformation potential leads to $T^4$-term (Eq. \ref{final7}) and
the unscreened deformation potential results in $T^2 \ln T $-term
(Eq. \ref{final5}).

The effects of the electron-electron interaction dominate in
conductivity of weakly disordered conductors at low temperatures.
At higher temperatures, conductivity is determined by the
electron-phonon-impurity interference and then the pure
electron-phonon scattering prevails over the interference
mechanisms. Relative values of interference terms and
characteristic crossover temperatures depend on many parameters.
The effects of the electron-electron interaction are enhanced in
low-dimensional conductors. As we discussed in Sec. IV, at low
temperatures the contribution of the electron-phonon-impurity
interference turns out to be $ (p_F/q_T)^2/(q_T l)$ times larger
than the contribution of pure electron-phonon scattering.
Therefore, the interference contributions dominate over pure
electron-phonon scattering at $T\leq u p_F (p_F l)^{-1/3}$. It is
important, that all interference corrections are proportional to
the Drude conductivity, and this characteristic feature may be
used for their experimental identification, as it has been done
for metallic films. \cite{Pt}

Note, that the $T^2$-term has been actually widely observed in
conductivity of doped semiconductors. It was associated with
strongly anisotropic Fermi surfaces and electron-electron
scattering (for a review see the book \cite{Gant}). In our
opinion, the electron-phonon-impurity interference correction is a
more plausible reason for such term.

\begin{acknowledgments}
The research was supported by the ONR grant. We would like to
thank I. Aleiner, J. Bird, M. Gershenson, B. Narozhny, and D.
Maslov for useful discussions.
\end{acknowledgments}

\appendix*
\section{Quantum Transport Equation}

The goal of this section is to show how the basic equations
discussed in Sec. II are obtained in the formalism of the quantum
transport equation. This method is based on the Keldysh
diagrammatic technique, where the electric current is expressed
through the kinetic (Keldysh) component of electron Green
function, $G^C({\bf p},\epsilon)$, in the following way
\begin{equation}\label{Je}
{\bf J}_e=\sigma {\bf E} = e \int {d{\bf p} d\epsilon \over
(2\pi)^4} \ {\bf v} \ \ {\rm Im}G^C({\bf p},\epsilon).
\end{equation}
Without interaction effects, the expressions for electron Green
functions in disordered conductors are well known. The retarded
(R) and advanced (A) electron Green functions are
\begin{eqnarray}\label{GR}
G^R_0({\bf p},\epsilon)= [G^A_0({\bf p},\epsilon)]^* =
(\epsilon-\xi_p+i/2\tau)^{-1},
\end{eqnarray}
and $\xi_p=(p^2-p^2_F)/2m$. The kinetic component of the electron
Green function is given by
\begin{eqnarray}\label{Gc}
G^C_0({\bf p},\epsilon)&=& 2i \ S({\bf p},\epsilon) \ \Im
[G^A_0({\bf p},\epsilon)] + \delta G^C_0, \\
 \delta G^C_0 &=& {1\over 2}\lbrace S_0 (\epsilon),
G^A_0+G^R_0\rbrace.
\end{eqnarray}
where the electron distribution function is given by
\begin{eqnarray}\label{F}
S({\bf p}, \epsilon) &=& S_0(\epsilon)+\phi({\bf p}, \epsilon),
\\ S_0(\epsilon)&=& -\tanh(\epsilon/2T), \\ \phi({\bf p},
\epsilon)&=& -e\tau ({\bf v}\cdot {\bf E}) {\partial
S_0(\epsilon)\over
\partial \epsilon}.
\end{eqnarray}
The nonlocal correction $\delta G^C_0({\bf p},\epsilon)$ has a
form of the Poisson bracket,
\begin{eqnarray}\label{E}
\lbrace A,B \rbrace =e {\bf E}  \ \biggr( {\partial A \over
\partial \epsilon } {\partial B \over \partial {\bf p} } -
{\partial B \over \partial \epsilon }
  {\partial A \over \partial {\bf p} } \biggl).
\end{eqnarray}
The kinetic Green function $G^C_0$(Eq. \ref{Gc}) with the
distribution function $S({\bf p}, \epsilon)$ (Eq. \ref{F}) takes
into account only elastic electron scattering from impurities.
Substituting $G^C_0$ in Eq. \ref{Je}, one could get the Drude
conductivity. The nonlocal correction $\delta G^C_0({\bf
p},\epsilon)$ is required to satisfy the unitary condition for the
matrix Keldysh function, \cite{LO} it removes divergence in Eq.
\ref{Je} far from the Fermi surface.

Assuming that the elastic electron scattering from impurities and
defects dominates in the electron momentum relaxation, one can
apply the iteration procedure to the Dyson equation. Then the
many-body correction to the kinetic Green function is expressed
through the electron self-energy in the following way \cite{RSLW},
\begin{eqnarray}\label{deltaGc}
 \delta_{int} G^C &=& 2 \tau \ [ \Sigma_{int}^C-2i \ S \ \Im
\Sigma_{int}^A ] \ \Im G^A_0 \nonumber
\\  &+& 2 \tau \ \lbrace \Re \Sigma_{int}^A, \ S_0\rbrace \
 \Im G^A_0  +  S \ \Im \ \Sigma^A_{int} (G^A_0)^2. \ \ \ \ \
\ \ \
\end{eqnarray}
Note, that the electric field enters Eq. \ref{deltaGc} through the
nonequilibium distribution function $\phi({\bf p}, \epsilon) $
(Eq. \ref{F}) and the electric Poisson bracket (Eq. \ref{E}), the
interaction is included in components of the electron self-energy.

For weakly disordered conductor we should consider three electron
self-energy diagrams shown in Fig. 1. The first diagram does not
consists of dotted lines corresponding to the electron-impurity
interaction. Electron-impurity scattering is included only in the
electron Green functions (Eq. \ref{GR}). Therefore, without
nonlocal quantum corrections in the form of the Poisson brackets
(Eq. \ref{E}), the first diagram results in the Bloch-Gruneisen
term. Interference effects are taken into account by the Poisson
bracket terms. Namely, $\delta C^C_0$ (Eq. \ref{Gc}) has to be
taken into consideration in all self-energy components (A,R,C) in
the first term in Eq. \ref{deltaGc}. The second term has the
Poisson bracket form and, therefore, it is directly calculated
with the equilibrium distribution functions. Substituting
$\delta_{int} C^C$ in Eq. \ref{Je}, we get the correction to the
conductivity in the form of Eqs. \ref{sigma} and \ref{eF} with the
function $\Phi(q, \omega)$ given by
\begin{eqnarray}\label{Phi1}
&&\Phi_1(q, \omega) = {i \ n \over 4 \pi l^2} \int d\xi_p \
\langle \langle \ \gamma^2 ({\bf v} \cdot {\bf e})^2 [G^{R} ({\bf
q}+{\bf p},\epsilon + \omega)]^2 \nonumber \\ &&\times G^A({\bf
p},\epsilon) \rangle_{\bf v} \rangle_{\bf q} = -{1\over 2 \tau^2
v_F^2} \biggl\langle \biggl\langle{\gamma^2 ({\bf v}_F \cdot {\bf
e})^2 \over ({\bf q}{\bf v}_F -\omega -i0)^2}\biggr\rangle_{\bf v}
\biggr\rangle_{\bf q}. \nonumber \\
\end{eqnarray}

In the second and third diagrams the electron-impurity and
electron-electron (electron-phonon) interactions are directly
presented. Therefore, interference contributions of these diagrams
originate from the first term in Eq. \ref{deltaGc} without any
Poisson bracket corrections. Calculations show that the
contribution of the second diagram is exactly the same as the
first one,
\begin{eqnarray}\label{Phi2}
\Phi_2(q, \omega)=\Phi_1(q, \omega).
\end{eqnarray}
The third diagram with the vertex $\gamma$ renormalized by elastic
electron-impurity scattering gives
\begin{eqnarray}\label{Phi3}
\Phi_3(q, \omega) = \langle K(q,\omega)\cdot L({\bf
q},\omega)\rangle_{\bf q}- \ \langle [M({\bf q},\omega)]^2
\rangle_{\bf q}. \nonumber \\
\end{eqnarray}
The first term in Eq. \ref{Phi3} corresponds to the third diagram
with the equilibrium vertex, which is given by
\begin{eqnarray}\label{K}
 K(q,\omega) &=&  {1\over  \pi \nu_n \tau} \int {d{\bf p} \over
(2\pi )^n} \gamma \ G^A({\bf p},\epsilon)  G^{R}({\bf p}+ {\bf
q},\epsilon +\omega) \nonumber \\ &=& -{i\over \tau} \biggl\langle
{\gamma \over {\bf q}{\bf v}_F -\omega -i0}\biggr\rangle_{\bf v},
\end{eqnarray}
and the rest of the diagram is described by the function
\begin{eqnarray}\label{L}
 L({\bf q},\omega) &=&  {1\over  \pi \nu_n \tau} \int {d{\bf p} \over
(2\pi )^n} \ G^A({\bf p},\epsilon)  G^{R}({\bf p}+ {\bf
q},\epsilon +\omega) \nonumber
\\ \times  {\gamma  ({\bf v} \cdot {\bf e})^2 \over v^2}
&=& -{i\over \tau v_F^2} \biggl\langle {\gamma ({\bf v}_F \cdot
{\bf e})^2 \over {\bf q}{\bf v}_F -\omega -i0}\biggr\rangle_{\bf
v}.
\end{eqnarray}
The nonequilibrium vertex calculated with the distribution
function $\phi({\bf p}, \epsilon) $ leads to
\begin{eqnarray}\label{M}
 M({\bf q},\omega) &=&  {1\over  \pi \nu_n \tau} \int {d{\bf p} \over
(2\pi )^n}  G^A({\bf p},\epsilon) G^{R}({\bf p}+ {\bf q},\epsilon
+\omega) \nonumber \\ \times {\gamma  {\bf v} \cdot {\bf e}\over
v} &=& -{i\over \tau v_F} \biggl\langle {\gamma \ {\bf v}_F \cdot
{\bf e}\over {\bf q}{\bf v}_F -\omega -i0} \biggr\rangle_{\bf v} .
\end{eqnarray}

Finally, summarizing $\Phi_i(q, \omega)$ (Eqs. \ref{Phi1},
\ref{Phi2}, and \ref{Phi3}), we get Eq. \ref{phi'}.

As we already mentioned, Eqs. \ref{sigma} and \ref{phi'} can be
also derived in the linear response formalism. \cite{RS1,Al} The
corresponding diagrams are shown in Fig. 2. Here, the diagrams 1
and 2 corresponds to the first diagram of the transport equation
method (Eq. \ref{Phi1}), the diagrams 3 corresponds to the second
diagram (Eq.  \ref{Phi2}), the diagrams 4 and 5 are equivalent to
the third diagram with the equilibrium vertex (Eqs \ref{Phi3} and
\ref{K}), and the diagrams 6 and 7 are equivalent to the third
diagram with the nonequilibrium vertex (Eqs \ref{Phi3} and
\ref{M}).
\newpage

\begin{figure}
\caption{Electron self-energy diagrams in the quasi-ballistic
limit. Wavy line stands for to the electron-electron or
electron-phonon scattering, a dotted line stands for to elastic
electron scattering from random potential, and a straight line
stands for the electron Green function.}

\end{figure}

\begin{figure}
\caption{Diagrams of the conductivity in the linear response
method.}

\end{figure}

\begin{figure}

\caption{Screening of the electron-phonon vertex. The zigzag line
stands for the Coulomb potential, $V(q,\omega)$.}

\end{figure}

\end{document}